\documentclass{article}
\usepackage{graphicx,amsmath,amsfonts,amssymb,amsxtra}
\usepackage{}

\newcommand{\lscooD}{La$_{5/3}$Sr$_{1/3}$CoO$_4$}
\newcommand{\lcoo}{La$_{2}$CoO$_4$}

\newcommand{\lsno}{La$_{2-x}$Sr$_{x}$NiO$_{4}$}
\newcommand{\lscooIV}{La$_{1.6}$Sr$_{0.4}$CoO$_4$}
\newcommand{\lsnoD}{La$_{5/3}$Sr$_{1/3}$NiO$_4$}
\newcommand{\coii}{Co$^{2+}$}
\newcommand{\coiii}{Co$^{3+}$}
\newcommand{\lscooV}{La$_{1.5}$Sr$_{0.5}$CoO$_4$}
\newcommand{\lsco}{La$_{2-x}$Sr$_{x}$CuO$_4$}
\newcommand{\lscoo}{La$_{2-x}$Sr$_{x}$CoO$_4$}

\newcommand{\FigNat}[1]{#1}

\title{Hour-glass magnetic excitations induced by nanoscopic phase separation in cobalt oxides La$_{2-x}$Sr$_x$CoO$_4$ \Large}

\date{}

\begin{document}
\maketitle
{Y. Drees$^{1,*}$, Z.~W.~Li$^{1,*}$, A. Ricci$^{2,*}$, M. Rotter$^{1}$, W. Schmidt$^{3}$,  D. Lamago$^{4}$, O. Sobolev$^{5,6}$, U. R\"{u}tt$^{2}$, O. Gutowski$^{2}$, M. Sprung$^{2}$, A. Piovano$^{7}$, J.~P.~Castellan$^{4}$,  \&  A.~C.~Komarek$^{1}$ }

\begin{enumerate}
\item Max-Planck-Institute for Chemical Physics of Solids, N\"othnitzer Str. 40, D-01187 Dresden, Germany
\item Deutsches Elektronen-Synchrotron DESY, Notkestr. 85, 22603 Hamburg, Germany
\item J\"{u}lich Centre for Neutron Science JCNS, Forschungszentrum J\"{u}lich GmbH, Outstation at ILL, BP 156, 6 Rue Jules Horowitz, 38042 Grenoble, France
\item Laboratoire L\'{e}on Brillouin, CEA/CNRS,F-91191 Gif-sur Yvette Cedex, UMR12 CEA-CNRS, B\^{a}t 563 CEA Saclay, France
\item Forschungsneutronenquelle Heinz Maier-Leibnitz (FRM-II), TU M\"{u}nchen, Lichtenbergstr. 1,  D-85747 Garching, Germany
\item Georg-August-Universit\"{a}t G\"{o}ttingen, Institut f\"{u}r Physikalische Chemie, Tammannstrasse 6, D-37077 G\"{o}ttingen, Germany
\item Institut Laue-Langevin (ILL), 6 Rue Jules Horowitz, F-38043 Grenoble, France\\
*: These authors contributed equally to this work
\end{enumerate}

\begin{abstract}
The magnetic excitations in the cuprate superconductors might be essential for an understanding of high-temperature superconductivity.
In these cuprate superconductors the magnetic excitation spectrum resembles an hour-glass and certain resonant magnetic excitations within are believed to be connected to the pairing mechanism which is corroborated by the observation of a universal linear scaling of superconducting gap and magnetic resonance energy.
So far, charge stripes are widely believed to be involved in the physics of hour-glass spectra.
Here we study an isostructural cobaltate that also exhibits an hour-glass magnetic spectrum.
Instead of the expected charge stripe order we observe nano phase separation and unravel a microscopically split origin of hour-glass spectra on the nano scale pointing to a connection between the magnetic resonance peak and the spin gap originating in islands of the antiferromagnetic parent insulator. Our findings open new ways to theories of magnetic excitations and superconductivity in cuprate superconductors.

\end{abstract}

\section{Introduction}

The spin excitations in high-temperature superconducting (HTSC) cuprates with a quasi two-dimensional layered crystal structure resemble an hour-glass if plotted as a function of energy and momentum transfer \cite{hourglassLSCO,hourglassLSCOb,hourglassLSCOc,hourglassLSCOd,hourglassLSCOe,hourglassLSCOf,hourglassLSCOg,hourglassLSCOh,hourglassLSCOi,pseudo,MagLSCOv,hourglass,drees}. Understanding the mechanism of these magnetic excitations and their relation to high-temperature superconductivity remains one of the great challenges in condensed matter physics.
Therefore, the recent discovery of such excitation spectra in an isostructural but insulating cobaltate reference system \lscooD\ has attracted enormous attention \cite{hourglass}.
The assumed presence of charge stripes in this cobalt oxide pointed to the presence of (fluctuating) stripes in the isostructural cuprates \cite{hourglass}
and, hence, suggested the importance of charge stripes for the physics in the high-temperature superconductors.
In charge stripe phases the doped charges (holes) segregate into stripes whereas antiferromagnetism is able to recover in the charge-depleted regions between these stripes.
However, the very recent finding of an hour-glass dispersion in a checkerboard charge ordered cobaltate \lscooIV\  has shown that hour-glass spectra are not necessarily connected to charge stripe phases \cite{drees}. \lscooD\ is a compound with a hole-concentration even more away from half-doping that exhibits diagonal magnetic satellite reflections around third-integer positions in reciprocal space. For this hole-doping the isostructural nickelates that also exhibit diagonal magnetic satellites at the same positions in reciprocal space exhibit the most stable diagonal charge stripe order. Hence, also for \lscooD\ the incommensurate magnetic peaks have been interpreted within a charge stripe scenario \cite{hourglass,musr}.
\par Here we demonstrate that there are no signatures of charge stripes in \lscooD.
Instead, we observe a fractal microstructure with a nanoscopic distribution of undoped and hole-doped regions indicating that
hour-glass spectra consist of excitations with distinct origin on the nanometer scale. Whereas the low-energy regime is still dominated by excitations within the nano-sized hole-doped regions, at highest energies only the nanoscopic undoped islands can be excited. In our nanoscale phase separation scenario the energy of the neck of the hour-glass could be interpreted as the spin-gap within undoped islands of type of the antiferromagnetic parent insulator \lcoo.

\section{Results}
\subsection{Hard X-ray diffraction measurements}
Using 100~keV hard X-rays at the synchrotron we first studied the charge correlations in \lscooD.
In a disordered charge stripe scenario the corresponding charge stripe ordering superstructure reflections should exhibit similar peak widths as the magnetic reflections \cite{andrade}.
But, in our highly precise synchrotron measurements we were not able to detect any indications for charge stripe ordering reflections (for L~=~0 and L~=~3~$\gg$~0)
in \lscooD\ with an unprecedented accuracy of about 10$^{-9}$ of a strong Bragg reflection, see \FigNat{Fig. 1 a-b}.
A charge stripe-ordered \lsnoD\ reference sample with same hole-concentration exhibits strong charge stripe ordering peaks in both equivalent scans.
Note, that the huge \coii- and the small \coiii-ions exhibit distinctly bigger differences in their ionic radii than the corresponding Ni-ions \cite{shannon} such that any occurring stripe phases should be much easier to detect in cobaltates than in nickelates. Furthermore, magnetic stripes and charge stripes can not be decoupled due to the nonmagnetic properties of the \coiii-ions \cite{half,chang} and due to the highly insulating localized nature \cite{chang} of this cobaltate system.
Instead of charge stripe ordering reflections we, surprisingly, observe weak checkerboard charge ordering intensities at half-integer $H$ and $K$ positions in reciprocal space that survive up to room-temperature, see \FigNat{Fig. 1 c-d}.
This important finding indicates that the robust checkerboard charge ordering correlations still persist also far away from half doping down to 1/3-doping in the \lscoo\ system and that there are no disordered charge stripe phases in this entire incommensurate magnetic regime of the \lscoo\ system ($1/3$~$\leq$~$x$~$<$~$1/2$).
Instead, checkerboard charge ordered regions with roughly 0.7(2)~nm correlation length appear.
The remaining \coii-ions have to be distributed somewhere into the cobalt oxygen planes since the hole-doping in \lscooD\ is far below half-doping. Hence, checkerboard charge ordered regions have to be interspersed with \lcoo-like undoped islands within a nano-phase separation scenario.

\subsection{Resonant micro X-ray diffraction measurements}
We analyzed the real space distribution of the charge correlations in \lscooD\ by means of resonant scanning micro X-ray diffraction ($\mu$XRD) measurements. The $\mu$XRD technique has
been already successfully applied for the study of phase separation in
the isostructural cuprates \cite{microA,microB}. Here, we additionally applied an incident X-ray energy of 7719~eV in order to measure any \coii/\coiii\ charge ordering correlations in resonance for a high contrast of charge ordering reflections. Moving the sample under a 2~$\mu$m focussed beam with an x-y translator we scanned a sample area of $202\times 210$~$\mu$m$^2$ collecting 10605 different diffraction patterns at $\sim$300~K. For each scanned point of the sample the (0.5~0.5~2) peak profile has been extracted and the full width half maximum (FWHM) along $ab$-direction has been evaluated in order to obtain the domain size of the checkerboard charge ordered regions. The spatial map of this domain size along ab-direction is shown in \FigNat{Fig. 1 e}. In order to quantify \cite{microC} the distribution of undoped islands and checkerboard charge ordered regions, we calculated the probability density function (PDF) of the measured domains sizes, see \FigNat{Fig. 1 f}. Similar as observed in the isostructural cuprates \cite{microA,microB} for oxygen ordering superstructure reflections the
PDF tail presents a perfect power-law behavior and can be fitted by the function $\text{PDF}(x) = x^{-\alpha}$, where $x$ is the $ab$-domain size of checkerboard charge ordering and $\alpha$~$=$~$1.99(8)$ is the critical exponent of the power-law. This particular behavior highlights a scale-free distribution of the checkerboard charge ordered domains size in \lscooD. From the non power-law region of the PDF we can estimate the expected values of the checkerboard charge ordering ab-domains size that amounts to $\sim$0.8(2)~nm. Thus, our $\mu$XRD measurements indicate a fractal microstructure up to a maximum observed domain size of $\sim$39~nm and corroborate the presence of a disordered checkerboard charge ordered phase interspersed with undoped islands and, hence, our nano phase separation scenario.
\\

\subsection{Neutron scattering measurements}
Studying the magnetic excitations in \lscooD\ in detail,
we surprisingly observe a hitherto unknown high-energy magnon band above the entire
classical hour-glass dispersion, see \FigNat{Fig. 2 a}.
The polarization analysis confirms that this additional high-energy mode is magnetic in origin.
Note, that this is the first observation of an additional separated high-energy magnon band appearing completely above an entire hour-glass dispersion.
Usually, one expects optical modes in a magnon dispersion without hour-glass shape or the optical magnon mode merges with the low-energy dispersion
such that both  together are forming an hour-glass-shaped dispersion \cite{optA,optB,optC}.
Also a 40\% Sr-doped cobaltate within the checkerboard charge ordered regime exhibits this same additional high-energy mode, see \FigNat{Fig. 2 c}.
Our polarized neutron measurements uncover the interesting nature of these magnetic excitations:
whereas all low-energy spin excitations within the entire classical hour-glass dispersion are mainly in-plane excitations (see \FigNat{Fig. 2 b}), the additional yet unknown high-energy mode appears to be an out-of-plane polarized magnon band, see \FigNat{Fig. 2 a}.
Since the low-energy excitations in \lscooD\ basically resemble on the excitations in the half-doped cobaltate \lscooV\ and since the high-energy excitations resemble on the excitations of the undoped parent compound \lcoo\ including the out-of-plane polarized additional high-energy mode, our observations suggest that the hour-glass dispersion is not one single dispersion, but, that it consists of two dispersions with distinct origin, see \FigNat{Fig. 3}. The first dispersion arises from the low-energy magnetic excitations of the hole-doped regions that are governed by frustration and that exhibit chiral or non-collinear magnetic phases and the second one arises from the high energy magnetic excitations of the undoped islands strongly resembling on the dispersion of the undoped parent compound (\lcoo) that exhibits a spin gap and an additional high-energy magnon mode. This model is shown schematically in \FigNat{Fig. 3}.
The intensity enhancement at the magnon merging point or neck of the hour-glass (we would like to call this by definition also simply resonance peak for the cobaltates throughout this article) that can be traced with energy scans at the planar antiferromagnetic (AFM) wavevector exhibits a surprisingly similar energy dependence like in equivalent scans for \lcoo, see \FigNat{Fig. 2 a}  (dark yellow dots). This clearly points to a related origin of this resonance peak energy in \lscooD\ and the spin gap in \lcoo. This is also consistent with the temperature dependence of this energy scan and a beginning of a closing of this spin gap with increasing temperature, see \FigNat{Fig. 4 a} and also compare Ref.~\cite{LaCoo} for the temperature dependence of the long range ordered undoped compound \lcoo\ (with higher magnetic ordering temperature).
On the other hand the high-energy part of the hour-glass-shaped dispersion in \lscooD\ does not exceed to exactly that high energies as the dispersion for pure \lcoo. This might be owed to the structural changes and the amount of disorder in the charge sector with extremely short-ranged \lcoo-like correlations. However, this difference in the maximum energy clearly proves that we did not simply measure the magnetic signal of a macroscopically phase separated sample with conventional phase separation.
Our polarized neutron study underlines our findings, since the hitherto unknown high-energy magnon band clearly belongs to an out-of-plane excitation whereas the outwards dispersing contributions from the neck of the hour-glass are mostly in-plane excitations, see \FigNat{Fig. 2 d}, and, thus, fully compatible to the \lcoo\ excitation spectrum \cite{LaCoo}.
Also the 40\% Sr-doped cobaltate exhibits qualitatively the same energy scan at the planar AFM wavevector, see \FigNat{Fig. 2 c}. This is fully in agreement with our scenario that the spin gap within the undoped islands is responsible for the resonance-like increase of intensity at the neck of the hour-glass, since the spin anisotropy gap energy should be much less doping dependent than a dispersion which starts from different incommensurate magnetic peak positions in samples with different hole-doping.
Also the comparison of magnetic intensities of high and low energy magnetic signal for two different hole-concentrations corroborates our whole scenario, since the additional high energy mode is strongest for low Sr-doping where larger undoped regimes can be expected, compare \FigNat{Fig. 2 e,f}.
Furthermore, the temperature dependence of the hour-glass dispersion is also consistent with our nano phase separation model, see \FigNat{Fig. 4}. The temperature dependence of the high energy excitations is different from the strongly temperature affected low energy part.
Fitting the magnetic peaks of the constant-E scans at 5~meV and at 29~meV (shown in \FigNat{Fig. 4 b,c}) with two symmetric gaussians indicates a different temperature dependence in both energy regions.
In our model, this could be explained by a different decrease of the spin wave stiffness of upper and lower dispersions with rising temperature, see  \FigNat{Fig. 4 d}. Furthermore, the peak widths are unchanged with temperature, see inset in \FigNat{Fig. 4 d}. This unaffected peak width reflects an almost temperature independent dynamic correlation length. In our nano phase separation model this would be naturally implied by the limiting factor of the limited static domain size of hole-doped checkerboard charge ordered and undoped islands. For the low-energy excitations this dynamic correlation length would be of the order of roughly 0.6~nm which would nicely correspond to the measured average correlation length of the static checkerboard charge order which amounts to roughly 0.7~nm.

\section{Discussion}

\par The physical reason for the occurrence of these two distinct sets of dispersions that we schematically present in \FigNat{Fig. 3 b}  can be understood as well. The magnetic exchange interaction $J'$ between two \coii-ions in the hole-doped (checkerboard charge ordered) regions is much smaller than the comparably large magnetic interactions $J$ within the undoped islands of \lcoo-type ($J\gg J'$), see \FigNat{Fig. 3 a}. Thus, the excitations within the hole-doped regions stay at low energies whereas the excitations of the undoped islands (red area in \FigNat{Fig. 3 a} ) exceed to high energies. The reason that the excitations within both types of regions are that strongly de-coupled is the huge difference in the exchange couplings $J$ and $J'$ that is of the order of $\sim$9.7~meV and $\sim$1.4~meV respectively \cite{LaCoo,half} and also the existence of a sizeable spin gap within the phase with larger $J$. For a simplified mechanical analogy see \FigNat{Fig. 3 c}.
The neck of the hour-glass dispersion results basically from the existence of a spin gap within the undoped islands. Note, that  such a spin gap of exactly these dimensions has been observed in \lcoo\ \cite{LaCoo}, see \FigNat{Fig. 2 a}  (dark yellow dots).

Our spin wave calculations based on our nano-phase separation scenario, that we modeled with Monte Carlo simulations using the Metropolis algorithm, are able to reproduce all basic features of the hour-glass dispersion including our additional high-energy mode and, thus, corroborate our experimentally found nano phase separation model, see Methods section and \FigNat{Fig. 5}.
First of all, chiral or non-collinear magnetic structures arise due to frustration in analogy to a proposed scenario for LSCO \cite{LSCO,LSCOb}.
More important, our calculations are able to reproduce the surprising additional high-energy mode that arises from out-of-plane excitations exactly as observed in our polarized neutron scattering experiments (see also the movie in the suppl. material). Furthermore, our simulations indicate - in close relation to our mechanical analogy shown in \FigNat{Fig. 3 c} - that the hole-doped (checkerboard charge ordered) regions can be only excited at lower energies whereas at high energies only the undoped islands of type of the antiferromagnetic parent insulator \lcoo\ can be excited, see \FigNat{Fig. 5}. Hence, our spin wave simulations corroborate our experimental finding of a microscopically split origin of upper and lower part of magnetic excitations within the so-called hour-glass dispersion.

\par Finally, the insulating cobaltates appear to be a copper-free reference system that enables us to understand some most basic underlying mechanisms of hour-glass spectra and that overcomes the obstacle of fluctuating charges, thus, unraveling a direct connection between the occurrence of hour-glass magnetic spectra and nano phase separated microstructures.
Our findings might help to understand the observations of isotropic high-energy magnetic excitations in the isostructural \lsco\ (LSCO) materials \cite{MagLSCOisotropic,MagLSCOv,hourglassLSCOg}
or the recent observations of similar high-energy magnetic excitations in superconducting LSCO samples and the undoped parent materials probed by resonant inelastic X-ray scattering (RIXS) \cite{RIXb,RIXa}. Whereas it seems to be reasonable to attribute the spin gap within the undoped islands of the cobaltates to an anisotropy gap like in \lcoo\ \cite{LaCoo},  in the cuprates the higher energy scales within the excitation spectra would suggest a different physics. Spin gaps with higher energies are known for singlet-triplet excitations of cuprate materials due to finite size effects of the undoped regions   \cite{spingapA,spingapB,spingapC,spingapD}.

\textbf{{Methods \Large}} \\
\subsection{Crystal synthesis}
The author A.~C.~K. has grown the \lscoo\ and \lsno\ (LSNO) single crystals by the optical floating zone technique using a CSC four-mirror image furnace following the route described in Ref.~\cite{drees}.
The crystals are single phase as ascertained by high-resolution powder X-ray diffraction measurements.
The single crystals are single domain as ascertained by Laue X-ray and single crystal neutron diffraction techniques.
Due to the measured incommensurability in these neutron measurements the oxygen content in La$_{5/3}$Sr$_{1/3}$CoO$_{4+\delta}$ is close to stoichiometric: $\delta\sim 0.01$.
The crystal quality was further ascertained by WDX measurements indicating close to nominal stoichiometry within our single crystals:
La$_{1.675(6)}$Sr$_{0.325(2)}$CoO$_{4+\delta}$ (averaged over 10 measurements).
Finally, we performed a highly accurate single crystal X-ray diffraction measurement at a Bruker D8 VENTURE single-crystal X-ray diffractometer equiped with a Photon large area CMOS detector. The sample has a spherical shape with 95(5)~$\mu$m diameter. In total, 93028 reflections have been collected using Mo K$_{\alpha}$ radiation and a bent graphite monochromator for $\sim$3 times intensity enhancement. The internal R-value amounts to 2.20\% with a redundancy of 50.4. The R and weighted R value in the refinement on $F^2$ amount to 1.66\% and 4.19\% and the goodness of fit amounts to 1.27. The resulting structural parameters within space group I4/mmm are $z$(La/Sr)= 0.362234(1), $U_{iso}$(La/Sr)=0.005885(2), $U_{iso}$(Co)=0.004560(4), $U_{iso}$(O1)=0.009831(25), $z$(O2)=0.173122(24), $U_{iso}$(O2)=0.020418(36), $a$~= 3.8552(3)~\AA, $c$~= 12.6267(9)~\AA. And the refinement of the Sr-occupancy indicates that $x$ amounts to 33.2(2)\% which is close to stoichiometric.
The nominally 40\% Sr doped reference sample was already described in Ref.~\cite{drees} and has a composition of La$_{1.591(3)}$Sr$_{0.404(1)}$CoO$_{4.01(1)}$ according to ICP-OES and titration measurements.

The co-author Z.~W.~L. has also grown another La$_{5/3}$Sr$_{1/3}$CoO$_{4+\delta}$ single crystal following the route described in Ref.~\cite{hourglass} which reveals larger additional oxygen contents than the sample grown by the author A.~C.~K. and was only used for the measurement of the temperature dependence of the magnetic excitations and the $\mu$XRD experiments. Comparing the magnetic incommensurabilities the excess of oxygen amounts to $\delta$~$=$~$0.035(1)$ in this sample. However, the dispersion matches reasonably well with that of Ref.~\cite{hourglass}, thus, indicating a similar incommensurability.
Complementary inductively coupled plasma optical emission spectroscopy (ICP-OES) and WDX measurements indicate close to nominal stoichiometry of that sample: La$_{1.675(11)}$Sr$_{0.326(3)}$CoO$_{4+\delta}$ and La$_{1.663(7)}$Sr$_{0.330(6)}$CoO$_{4+\delta}$ respectively.

\subsection{Synchrotron measurements}
The synchrotron radiation single crystal X-ray diffraction measurements have been performed at beamline P07 at PETRA III at DESY
using 100~keV hard X-rays monochromatized by a bent Si(111) double monochromator providing an energy resolution of 0.2\%. The signal was analyzed with a Si-Ge(111) gradient crystal analyzer and detected with a point detector. The measured samples were very roughly (0.5-1.5)$^3$~mm$^3$ sized single crystals. The monitor-normalized $(200)$ reflection intensities of \lscooD\ and the LSNO reference sample amount to 2.7$\cdot 10^{10}$ photons/s and 9.7$\cdot 10^{9}$ photons/s.

The $\mu$XRD experiments were carried out at the Coherence Beamline P10 at PETRA III. The X-ray beam is monochromatized using a cryogenically cooled Si(111) double crystal monochromator. An X-ray energy of 7719~eV with a bandwidth of dE/E$\sim1.4\cdot10^{-4}$ was selected. About 1~m upstream of the focusing optics situated at 85.6~m from the undulator source, a slit of a size of 150$\times$75~$\mu$m$^2$ selected a highly coherent fraction of the X-ray beam. This collimated coherent X-ray beam was focused using the in-vacuum Beryllium compound refractive lens (CRL) transfocator of the P10 beamline to a size of about 2$\times$2~$\mu$m$^2$ at the sample position 2.2~m downstream of the transfocator center. A pair of JJ X-ray in vacuum slits located at 220mm and 800mm upstream of the sample were used to clean up the small angle scattering signal of the CRL's from focused X-ray beam. The incident flux on the sample was about 1-2$\times$$10^{11}$~photons/s. The scattered signal from the sample was recorded at a sample to detector distance of 25~cm using a Pilatus~300K detector.

\subsection{Neutron scattering measurements}
For the elastic and inelastic neutron scattering experiments one large \lscooD, \lscooIV\ or LSNO single crystal has been mounted with an [100]/[010] orientation in the scattering plane.
The inelastic neutron scattering experiments at the 1T.1 and 2T spectrometers have been performed in constant-k$_\text{f}$ operation mode with fixed focusing pyrolytic graphite (PG) monochromator and analyzer using two PG filters for the suppression of higher order contamination.
At the IN8 spectrometer an experimental setup with double focusing PG monochromator and analyzer as well as two PG filters was used.
At the IN22 spectrometer a setup with fixed vertically focusing (125 cm radius) and flat horizontal (111) Heusler monochromator and vertically focusing and flat horizontal (111) Heusler analyzer was used. For polarization analysis a CRYOPAD polarimeter for spherical polarization analysis was installed. The flipping ratio amounts to $\sim$16. A PG filter was used for suppression of higher order contamination.
At the IN12 spectrometer that is situated at the end of a horizontal elliptic focusing guide a double focusing PG monochromator and
a horizontally focusing Heusler analyzer were used. The incident neutron beam was polarised with a polarising cavity.
For polarization analysis a setup with Helmholtz coils for linear polarization analysis was installed. The flipping ratio amounts to $\sim$17.4. 
A velocity selector was used for elimination of higher order contributions (suppression better than 10$^{-4}$).
The temperature dependence of the magnetic excitations in La$_{5/3}$Sr$_{1/3}$CoO$_{4+\delta}$ was measured at the PUMA spectrometer using a double focusing PG monochromator and analyzer as well as two PG filters for elimination of higher order contamination.

\subsection{Numerical simulations}
The Monte-Carlo simulations for simulating disordered checkerboard charge ordered structures are based on the Metropolis algorithm similar as in Ref.~\cite{andrade}. Only one potential for Coulomb repulsion and another for steric repulsion of two adjacent \coii-ions with large ionic radii has been considered for our simulation of the charge distribution on a $N\times N$ lattice ($N$~$=$~30, 50) and statistical/random distribution of the charges at each site with a certain probability (here 40\%) in the initial step for about a dozen configurations.

In order to model the magnetism we introduce AFM
exchange interactions between the Co$^{2+}$ ions $i,j$ with
spin operators $  \mathbf{\hat S} ^i$ and $ \mathbf{\hat S} ^j$, respectively
(the Co$^{3+}$ ions were treated as nonmagnetic):
$
H_{ex}=-\frac{1}{2}\sum_{i,j} J(ij)  \mathbf{\hat S}^i  \mathbf{\hat S}^j
$.
According to our nano-phase separation model, we only assume
short range interactions J(100)=-5.8~ meV and J(200)=-0.85~meV.
In order to simplify the numerical calculation we consider only the Kramer's
ground state doublet ($|\pm>$) of each Co$^{2+}$ ion.
For the easy plane anisotropy of each Co$^{2+}$ ion
we assume an in plane magnetic saturation moment of
$
m_x=m_y=2\mu_B\langle-|\hat S_x|+\rangle=2i\mu_B\langle-|\hat S_y|+\rangle=3~\mu_B
$
and an out of plane saturation moment of
$
m_z=2\mu_B\langle+|\hat S_z|+\rangle =1.3~\mu_B
$. Thus, the exchange interaction $H_{ex}$ can be projected onto the following
spin--$1/2$ Hamiltonian, which is of the same form as that used in
Ref.~\cite{andrade}: $H=-\frac{1}{2}\sum_{i,j} J(ij)\sum_{\alpha=x,y,z}\frac{m_{\alpha}^2}{\mu_B^2} {\hat  s}_{\alpha}^i {\hat  s}_{\alpha}^j$.
Some representative $N \times N$ configurations with $N$~$=$~$30$ and $N$~$=$~$50$ were taken as the magnetic unit cell in a periodic
lattice and the spin structure and spin dynamics were calculated by the
Dynamical Matrix diagonalization method \cite{rotter} using the
McPhase program package \cite{mcphase}.
Finally, the spectra of all calculated configurations have been averaged.

\newpage

\bibliography{LaSrCoO}

 \par
\textbf{Acknowledgement} We thank O.~Stockert, C.~Sch\"{u}ssler-Langeheine, P. Thalmeier and L.~H.~Tjeng for helpful discussions.
       We thank A.~Todorova for assistance.
       We thank L.P.~Regnault for help at the IN22 spectrometer.
       For sample-characterization using powder X-ray diffraction and EDX/WDX measurements we thank H. Borrmann, Y. Prots and S. H\"{u}ckmann as well as the team of U.~Burkhardt respectively. \\

       \par
  \textbf{Contributions}  A.~C.~K. has conceived all experiments, performed the single crystal X-ray diffraction experiments, performed the Monte-Carlo simulations, analyzed the results and written the article.
      A.~C.~K. and Z.~W.~L. synthesized all studied materials.
      A.~C.~K., Y.~D., Z.~W.~L., O.~G. and U.~R. have performed the hard X-ray experiments.
      A.R., Z.~W.~L. and M.S. have performed the $\mu$XRD measurements.
      Y.~D., Z.~W.~L., W.~S., D.~L., J.~P.~C. and A.P. have measured the novel high-energy magnetic excitations with unpolarized and polarized neutrons.
      A.~C.~K.,  Z.~W.~L. and O.~S. have performed the temperature dependent neutron scattering experiments.
      M.~R. and A.~C.~K. have performed the spinwave calculations. \\

\par
   The authors declare that they have no competing financial interests.\\
   \par
   \textbf{Correspondence should be addressed to A.~C.~K.~(email: Komarek@cpfs.mpg.de).}

 \begin{figure}
\begin{center}
\includegraphics*[width=0.75\textwidth]{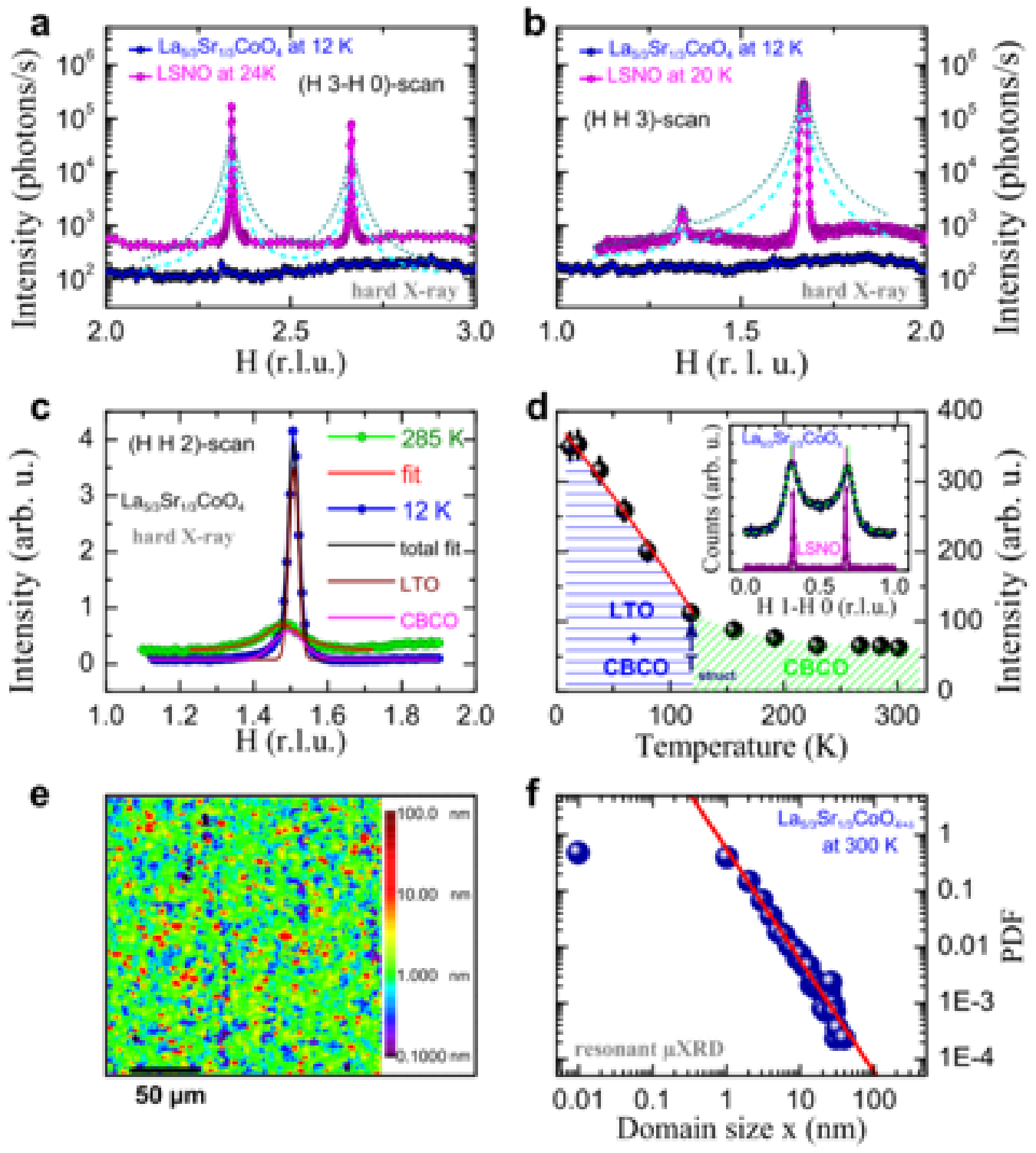}
\end{center}
\caption{\textbf{Charge correlations}   Synchrotron radiation single crystal X-ray diffraction measurements using 100~keV hard X-rays at beamline P07 at PETRA III. \textbf{a} Measured intensities of scans along ($H$~$3$$-$$H$~$0$) and \textbf{b} scans along ($H$~$H$~$3$) for \lscooD\ and a $\sim$1/3 hole-doped LSNO reference sample. Strong fundamental Bragg reflections are of the order of 10$^{10}$ photons/s. Intensities in \textbf{a-b} have been normalized to the monitor (incident beam) for a comparison of cobaltate and nickelate reference sample. The dashed lines are Lorentzian peaks with same integrated intensity as for the charge stripe ordering peaks of the LSNO reference sample but with peak widths that have been multiplied by the width ratio of the incommensurate magnetic peaks of both compounds measured with neutrons (inset of \textbf{d});  dotted lines are additionally normalized to the $(200)$ peak intensities. No diagonal charge stripe correlations like in the nickelates could be observed for \lscooD. \textbf{c} Weak checkerboard charge ordering reflections appear at half-integer positions for even values of $L$ already at room-temperature. The correlation length that could be obtained from Lorentzian fits indicates an average nanometer size of checkerboard charge ordered regions ($\sim$0.7(2)~nm). Below the structural transition tetragonal$\rightarrow$orthorhombic additional sharp octahedral tilting reflections appear on top of the broad checkerboard charge ordering reflections. \textbf{d} The temperature dependence of the total integrated intensities at half-integer peak positions. The inset shows neutron scans across the magnetic satellites for \lscooD\ and $\sim$1/3 hole-doped LSNO.
\textbf{e} Spatial map of the checkerboard charge ordered domain size distribution on the crystal surface along the $ab$-direction obtained from $\mu$XRD measurements at beamline P10 at PETRA-III. The distribution appears to be inhomogeneous and checkerboard charge-ordered regions with varying domain size can be observed.
\textbf{f} The PDF \cite{barabasi} of the checkerboard charge ordered domain size along the $ab$-direction. The behavior of the PDF tail obeys a perfect power-law ($\text{PDF}(x) = x^{-\alpha}$)  indicating a scale-free distribution in the checkerboard charge ordered domain sizes, thus corroborating our nano-phase-separation scenario. Only $\sim$1\% ($\sim$11\%) of the measured sample exhibits domain sizes bigger than 50 (10) unit cells. Error bars are s.d.
}
\label{fig1}
\end{figure}

\begin{figure}
\begin{center}
\includegraphics*[width=0.8\textwidth]{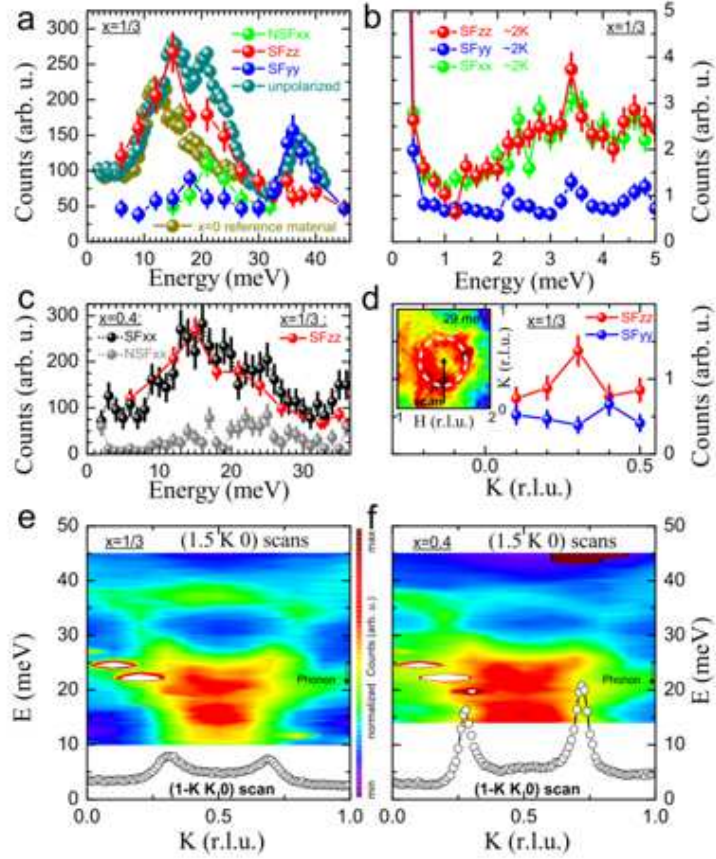}
\end{center}
\caption{\textbf{Spin excitations}  \textbf{a} Energy scans at the planar AFM wavevector studied by polarized neutrons. In-plane polarized excitations that can be detected in the Spin-Flip (SF) channel $||z$ ($z\perp$ scattering plane) emerge at the neck of the hour-glass (red data points). Surprisingly, an additional high-energy mode appears at elevated energies. The polarization analysis indicates the out-of-plane character of these high-energy excitations since it can be detected in the SF channel $||y$ (blue data points; $y$ within scattering plane, $y\perp\mathbf{Q}$). Additionally, also the results of unpolarized neutron scattering experiments of \lscooD\ measured at the IN8 spectrometer (dark cyan data points) and \lcoo\ taken from Ref.~\cite{LaCoo} (dark yellow data points) are shown. As can be seen, the small additional 20~meV feature belongs to a phonon whereas the resonance-like increase of intensity is located around 15~meV which is close to the in-plane anisotropy gap of \lcoo. \textbf{b} Study of the nature of the low-energy excitations measured by polarized neutrons measured by constant-Q scans at ($1/2$-$\varepsilon$~$1/2$+$\varepsilon$~0) at $\sim$2~K at the IN12 spectrometer. The in-plane character of these low-energy magnetic excitations is revealed. \textbf{c} Also \lscooIV\  exhibits the same energy scan at the planar AFM wavevector measured with polarized neutrons in the channel $||\mathbf{Q}$ that detects in-plane and out-of-plane excitations simultaneously. \textbf{d} For the outwards dispersing branches above the neck of the hour-glass also in-plane excitations can be detected in the SF channel $||z$. The inset shows the corresponding scan in reciprocal space within a 29~meV constant-E map. \textbf{e,f} Comparison of strength of the high-energy magnetic intensities for different Sr-doping measured at the IN8 spectrometer. Intensities have been normalized to the Phonon peak intensity (black arrrow). Intensity error bars are statistical error bars calculated by the square root of intensity.
}
\label{fig2}
\end{figure}

\begin{figure}
\begin{center}
\includegraphics*[width=1\textwidth]{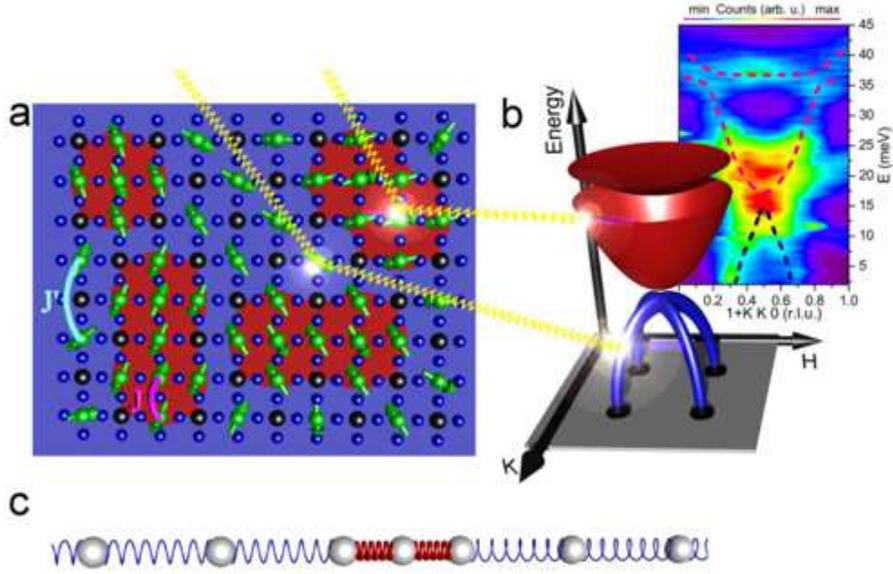}
\end{center}
\caption{\textbf{The two hour-glass dispersions}\ \textbf{a} Our nano-phase separation model within \lscooD. The doping of additional electrons or \coii-ions (green spheres) at \coiii-sites (black spheres) of an ideal checkerboard charge ordered structure of the half-doped AFM parent compound induces strong frustration as we already pointed out in Ref.~\cite{drees}. But due to the large ratio of the exchange interactions $J$/$J'$ the non-collinearity of the spin structure can be expected basically in the hole-doped regions (blue) and not within the undoped islands (red) where the spins should be aligned in a much more collinear manner. Neutrons (yellow waves) scattered from these different regions will behave differently.
\textbf{b} Schematic presentation of the magnetic spectra in \lscoo\ together with measured neutron scattering intensity behind. The blue contributions indicate the low energy excitations originating from the hole-doped regions (blue areas in \textbf{a}) with small $J'$ whereas the high-energy signal arises from excitations within the undoped regions (red areas in \textbf{a}) with large $J$. Here, the upper dispersion (red) is shifted somewhat to higher energies such that the lower dispersion (blue) becomes clearly visible although both dispersions should inter-penetrate each other. The yellow waves indicate neutrons that were either scattered from the undoped islands or from the hole-doped regions. \textbf{c} One-dimensional mechanical analogy of decoupled oscillations of balls within a one-dimensional chain that are connected with weak (blue) or strong (red) force constants.
}
\label{fig3}
\end{figure}

\begin{figure}
\begin{center}
\includegraphics*[width=0.8\textwidth]{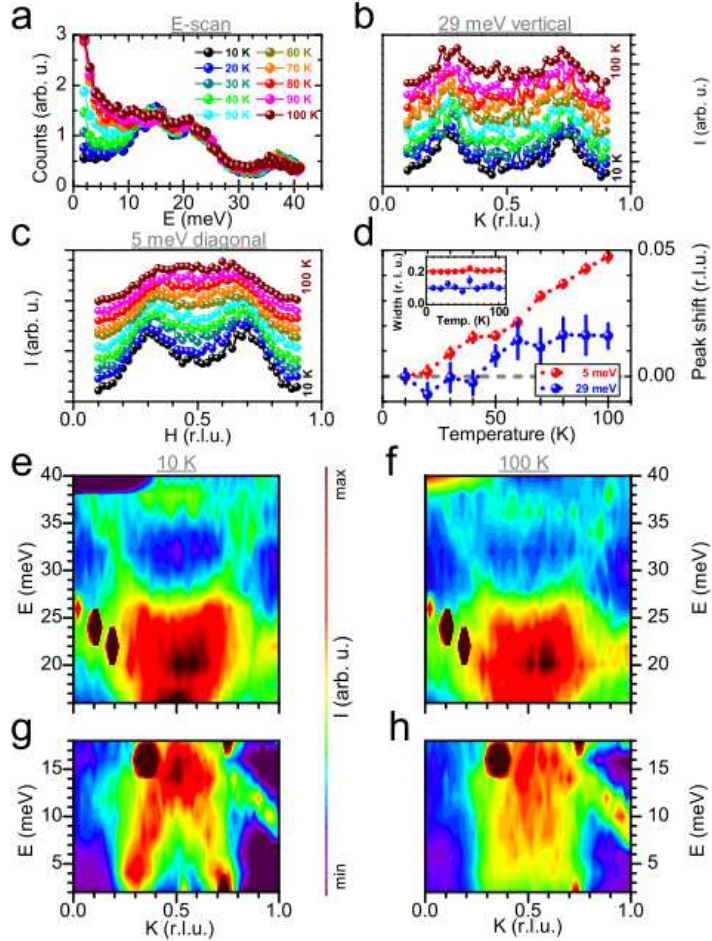}
\end{center}
\caption{\textbf{Temperature dependence of the hour-glass dispersion}\ \textbf{a} The temperature dependence of energy scans at the planar AFM wavevector.
\textbf{b} Temperature dependence of constant-E scans at 29~meV scanning vertical through the planar AFM wavevector.
\textbf{c} Temperature dependence of constant-E scans at 5~meV scanning diagonal through the magnetic satellite positions.
\textbf{d} Temperature dependence of the magnetic peak positions relative to the peak positions at lowest temperature (10~K) if two Gaussians are fitted to the data in figures \textbf{\emph{b,c}}. The inset shows the corresponding peak widths.
\textbf{e-h} Colour-contour plots of neutron scattering intensity measured as a function of energy and momentum transfer for low (10~K: \textbf{e,g}) and higher temperatures (100~K: \textbf{f,h}).  This temperature dependence is consistent with our nano phase separation model (see text). Intensities in \textbf{b-h} are multiplied by $k_{\text{i}}/k_{\text{f}} \cdot (1-\text{exp}(-\hbar\cdot\omega/k_{\text{B}}\cdot T))/f^2_\text{Co}(\mathbf{Q})$.  Error bars are s.d. (obtained from gaussian fits). Intensity error bars are statistical error bars calculated by the square root of intensity.
}
\label{fig3}
\end{figure}

\begin{figure}
\begin{center}
\includegraphics*[width=0.8\textwidth]{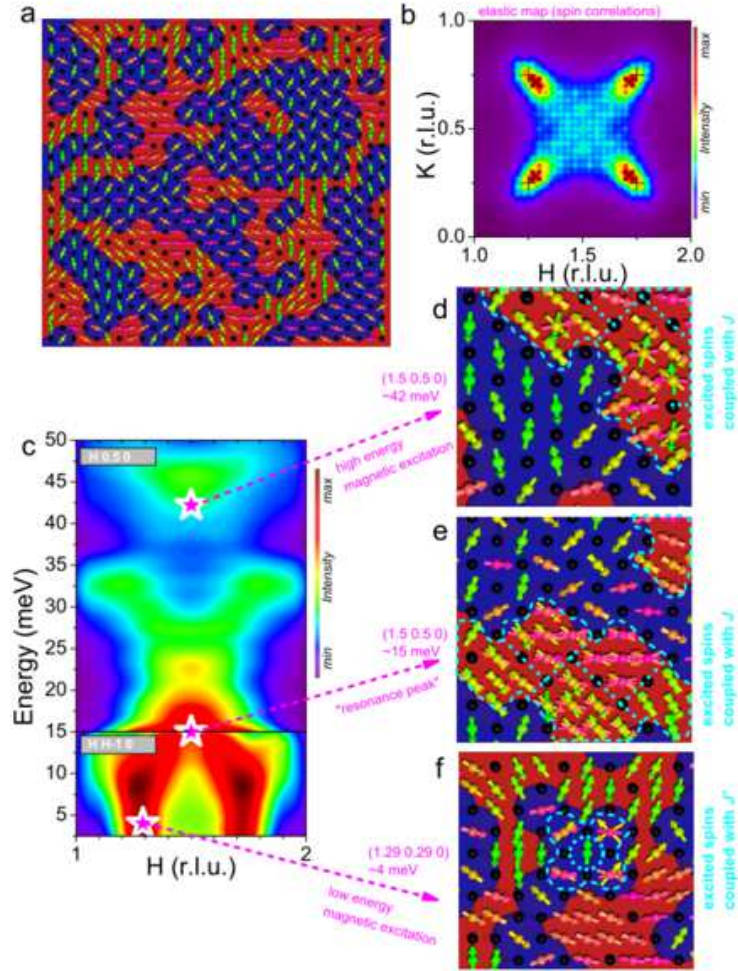}
\end{center}
\caption{\textbf{Spin wave simulations}\
\textbf{a} Spin correlations within a 30~$\times$~30 lattice calculated for our nano phase separation model. A non-collinear magnetic structure arises from frustration. Black spheres: nonmagnetic \coiii-ions, coloured spheres: \coii-ions, red shaded areas: undoped islands (\coii-ions coupled with at least one $J$), blue shaded areas: checkerboard charge ordered islands (\coii-ions coupled only with $J'$). The direction of spins is additionally indicated by a colour coding.
\textbf{b} Fourier transformation of the non-excited spin configurations in Q-space. The vertical crosses indicate wavevectors expected for an ideal checkerboard charge ordered system and the diagonal crosses indicate the expected incommensurate magnetic peak positions for our nominal hole-concentration according to an experimentally observed linear hole-doping dependence \cite{drees}.
\textbf{c} Simulated spin excitation spectrum. The basic features of the hour-glass dispersion can be reproduced.
\textbf{d-f} Calculated magnetic excitations. The movement of the spins is indicated by plotting  strongest displacements of the spins for this spin excitation on top of the static spin configuration such that direction and colour of that spin are differing when spins are involved in the excitation. Additionally, the dashed cyan lines mark regions of excited spins.
Our theoretical simulations yield a purely out-of-plane character of the novel high-energy spin excitations, see also our supplementary video file; (here in \textbf{d}, the out-of-plane oscillation has been artificially transformed into an in-plane excitation).
As can be seen, the checkerboard charge ordered regions (coupled with $J'$~$\ll$~$J$) can be excited basically only at low energies whereas at high energies only the undoped islands (coupled with $J$~$\gg$~$J'$) can be excited.
}
\label{fig3}
\end{figure}

\end{document}